\documentclass[12pt]{article}
\usepackage{jheppub}
\usepackage{amsmath,amssymb}

\makeatletter
\newcommand\fake@math{}
\def\fake@math#1\){{#1}}
\makeatother

\numberwithin{equation}{section}

\def\mop#1{\mathop{\rm #1}\nolimits}
\def\coth{\mop{coth}}

\def\Re{\mop{Re}}

\def\sgn{\mop{sgn}}

\title{A note on the SYK model with complex fermions}
\author[a]{Ksenia Bulycheva \footnote{On leave of absence from IITP, Bolshoy Karetny per. 19, Moscow, 127051 Russia}}

\affiliation[a]{Department of Physics, Princeton University, Princeton, NJ 08544}
\emailAdd{kseniab@princeton.edu}

\abstract{We consider a version of the Sachdev--Ye--Kitaev model with complex fermions. 
We apply the shadow formalism to find four-point functions in the leading order in $1/N$ and dimensions of operators present in the theory.
We also compute the retarded kernel and show that the Lyapunov exponent for the mode corresponding to the $U(1)$ charge is zero.}

\date{}
\begin{document}

\maketitle
\flushbottom

\tableofcontents

\section{Introduction}
The Sachdev--Ye--Kitaev model \cite{Sachdev:1992fk}, \cite{Kitaev:2015} and its various generalizations have received much attention in the recent years.
The original model proposed in \cite{Kitaev:2015} consisted of a large number of Majorana fermions in one spacetime dimension with random Gaussian four-fermion coupling. 
This model is remarkable for several reasons.
It flows to a (nearly) conformal theory in the infrared.
The full conformal symmetry appears to be spontaneously and explicitly broken, giving rise to a pseudo-Goldstone mode.
The presence of this mode causes the exponential growth of out of time order correlators, saturating the chaos bound of \cite{Maldacena:2015waa}.

This theory is also supposed to have a holographic dual \cite{Sachdev:2010um}.
Since SYK is a (nearly) $CFT_1$, it is naturally assumed to be dual to (nearly) $AdS_2$ gravity \cite{Almheiri:2014cka}.
Indeed, at low temperatures the effective action of the model is given by a Schwarzian derivative \cite{Maldacena:2016hyu}, which also appears in dilaton gravity in $AdS_2$ \cite{Jensen:2016pah}, \cite{Maldacena:2016upp}, \cite{Engelsoy:2016xyb}.
However, full understanding of a gravity dual of the SYK model is still lacking.

There are several generalizations of this model preserving these remarkable features.
These are two-dimensional versions of the model, formulated both on the lattice \cite{Gu:2016oyy} and in the continuum \cite{Turiaci:2017zwd}, models having flavor symmetry \cite{Gross:2016kjj} and supersymmetry \cite{Fu:2016vas}. 
There are also tensor models with non-random coupling sharing many properties of the SYK \cite{Witten:2016iux}, \cite{Klebanov:2016xxf}.
We focus on the SYK model with complex fermions.
It has been studied in \cite{Davison:2016ngz} from a thermodynamical perspective to compute transport coefficients of a strange metal.

The generalization we are studying here is the SYK model with global $U(1)$ symmetry. 
In addition to the pseudo-Goldstone mode of the original SYK, or $h=2$ mode of \cite{Maldacena:2016hyu}, the complex SYK model contains a mode associated with $U(1)$ charge. 
Since the $U(1)$ charge is conserved, the corresponding mode has the dimension $h=0$.
The symmetry of the spectrum under $h \leftrightarrow 1-h$ identifies it with a mode of positive dimension, $h=1$.

This $h=1$ mode is in some ways similar to the $h=2$ mode. 
In the real SYK model, the four-point function is a sum over eigenfunctions with integer $h$, including the $h=2$ mode. 
But since this mode corresponds to an existing operator in the spectrum, it makes the four-point function to diverge.
This divergence is later cured with $\left( \beta J \right)^{-1}$ corrections.
The same mode also contributes to the Lyapunov exponent of the out of time order correlators.

One might expect that the $h=1$ mode in the complex SYK also causes divergence in the four-point function and exhibits chaotic behaviour.
In this note, we repeat the analysis of \cite{Maldacena:2016hyu} for complex fermions.
In Section \ref{sec:model} we formulate the model and give the outline of the calculation.
Our goal is to find the four-point function of the model with complex fermions, which is an eigenfunction of the conformal Casimir.
In Section \ref{sec:symmetries} we discuss the discrete symmetries of the four-point function.
The first discrete symmetry we are interested in is time-reversal $\mathcal T$.
In general, the eigenfunctions of the Casimir may or may not be invariant under $\mathcal T$.
For $q=4n$, the SYK Hamiltonian is real and therefore preserves $\mathcal T$; however for $q=4n+2$ it is imaginary and $\mathcal T$-violating.
But since ladder diagrams contain an even number of couplings, the four-point functions of the SYK model are nevertheless $\mathcal T$-invariant.
So we focus on the $\mathcal T$-invariant four-point functions, although the $\mathcal T$-violating states may still be relevant in next orders in $1/N$ expansion.

Likewise, the eigenfunctions can be either odd or even under the $\chi \to \frac{\chi}{\chi-1}$ symmetry; those which are even under both this symmetry and time-reversal form a basis for four-point functions of the original SYK model with real fermions.
To include complex fermions, we should also consider eigenfunctions which are odd under $\chi \to \frac{\chi}{\chi-1}$.
In Section \ref{sec:shadow} we construct these eigenfunctions and write an explicit expression for the four-point functions, using the eigenvalues of the SYK kernel from Section \ref{sec:kernel}.
We see that the four-point function has a pole at $h=1$.
In Section \ref{sec:h=1} we study the $h=1$ mode in the solvable case of large $q$ and find that the pole is removed by inverse coupling corrections. 
Finally in Section \ref{sec:chaos} we analytically continue to real time and study chaotic behavior of the four-point function.
We find that although the mode corresponding to the $U(1)$ charge is an eigenfunction of the retarded kernel and therefore can contribute to the chaos, its Lyapunov exponent is zero.

{\bf Acknowledgements.} The author is grateful to Edward Witten for suggesting the problem and numerous discussions and also to Grisha Tarnopolsky for many productive conversations. 
While finishing the draft, the author noticed that the paper \cite{Peng:2017spg} has considerable overlap with the analysis presented here.

\section{Complex SYK model}
\label{sec:model}
We consider a one-dimensional system of $N$ complex fermions with $q$-particle random interaction ($\bar{\psi}$ denotes complex conjugate):

\begin{equation}
    H=i^{\frac{q}{2}}\sum j_{\bar{\imath}_1 \dots \bar{\imath}_{\frac{q}{2}} {i}_{\frac{q}{2}+1}\dots {i}_q} \bar{\psi}_{\bar{\imath}_1}\dots\bar{\psi}_{\bar{\imath}_{\frac{q}{2}}}\psi_{i_{\frac{q}{2}+1}} \dots \psi_{{i}_q}, \qquad 1\le i_k \le N,
    \label{H_def}
\end{equation}
with real gaussian coupling $j$ scaling with $N$:

\begin{equation}
    \langle  j_{\bar{\imath}_1 \dots \bar{\imath}_{\frac{q}{2}} {i}_{\frac{q}{2}+1}\dots {i}_q}^2 \rangle=\frac{J^2 \left( q-1 \right)^2}{N^{q-1}},
    \label{j_mean}
\end{equation}
and $N$ taken to be large.
This theory is a straightforward generalization of the Kitaev's model \cite{Kitaev:2015}, and by the same token it has a (near) conformal limit at large coupling $\beta J \gg 1$.

%
%
We expect the complex model to have the same pseudo-Goldstone $h=2$ mode as the real SYK, corresponding to the operator:

\begin{equation}
    \mathcal O_2=\bar{\psi}_{i}\partial_t \psi_i.
    \label{O2_def}
\end{equation}

In addition to that, we expect it to have the $U(1)$ charge operator. 
Since the $U(1)$ symmetry $\psi \to e^{i\alpha}\psi$ is conserved, it should have conformal dimension zero:

\begin{equation}
    \mathcal O_0=\bar{\psi}_i\psi_i.
    \label{O0_def}
\end{equation}

In what follows, we mostly work in Euclidean time and at zero temperature, until in Section \ref{sec:chaos} we discuss analytic continuation to real time and out of time order correlators. 

\begin{figure}
    \centering
    \psfig{file=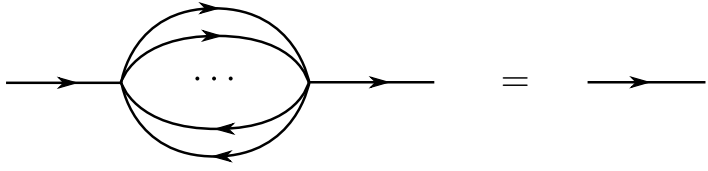, width=.6 \textwidth}
    \caption{Schwinder-Dyson equation for the two-point function. Black line stands for the full conformal propagator.}
    \label{fig:SD_eq}
\end{figure}

In the large $N$ limit, the SYK model  is dominated by melonic graphs. 
This allows us to find correlators using functional methods.
The two-point function obeys the Schwinger--Dyson equation, reflecting the fact that the leading correction to the propagator comes from inserting a ``melon'' (see fig. \ref{SD_eq}):

\begin{equation}
    \int d\tau' J^2 G\left( \tau_1,\tau' \right) G^{q-1}\left( \tau',\tau_2 \right)=-\delta\left( \tau_1-\tau_2 \right).
    \label{SD_eq}
\end{equation}

Solving this, one can find the propagator \cite{Polchinski:2016xgd}:

\begin{equation}
    \langle \bar{\psi}_{i}\left( \tau_1 \right) \psi_j\left( \tau_2 \right)\rangle = \delta_{ij}  G\left( \tau_1,\tau_2 \right) = \delta_{ij}\frac{b \sgn \left( \tau_1-\tau_2 \right)}{|\tau_1-\tau_2|^{2\Delta}},
    \label{G_def}
\end{equation}
where
\begin{equation}
    \Delta=\frac{1}{q}, \qquad J^2 b^q \pi=\left( \frac{1}{2}-\Delta \right)\tan \pi \Delta.
    \label{b_def}
\end{equation}

\begin{figure}
    \centering
    \psfig{file=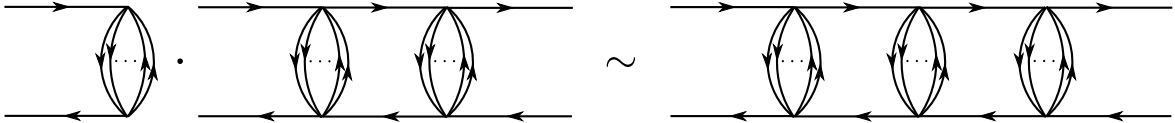, width=\textwidth}
    \caption{Kernel adds a rung to the ladder diagram.}
    \label{fig:kernel}
\end{figure}

The next step is to find the four-point function of the model,
\begin{equation}
    \left\langle \bar{\psi}_i\left( \tau_1 \right)\psi_i\left( \tau_2 \right)\bar{\psi}_j\left( \tau_3 \right)\psi_j\left( \tau_4 \right)\right\rangle.
    \label{4pt_ij}
\end{equation}
For $i\neq j$ and in the leading order in $N$, this is just a product of propagators. 
The leading $1/N$ correction comes from the ladder diagrams, as in the fig. \ref{fig:kernel}:

\begin{equation}
    \langle \bar{\psi}_i\left( \tau_1 \right) \psi_i\left( \tau_2 \right) \bar{\psi}_j\left( \tau_3 \right) \psi_j\left( \tau_4 \right) \rangle=G\left( \tau_1,\tau_2 \right)G\left( \tau_3,\tau_4 \right)+\frac{1}{N}\mathcal F\left( \tau_1,\tau_2,\tau_3,\tau_4 \right)+O\left( \frac{1}{N} \right).
    \label{F4pt_def}
\end{equation}
The correction $\mathcal F$ is an infinite sum of ladder diagrams with different numbers of ``rungs'':

\begin{equation}
    \mathcal F = \mathcal F_0 +\mathcal F_1 + \mathcal F_2 + \dots. 
    \label{F_def}
\end{equation}
Adding a rung to the ladder can be represented by acting with a differential operator, or ``kernel'', on the ladder diagram with a given number of rungs:

\begin{equation}
    K \circ \mathcal F_i=\mathcal F_{i+1},
    \label{KF_def}
\end{equation}

We define $K$ precisely later in Section \ref{sec:kernel}.
Then the full four point function becomes a sum of a geometric progression:
\begin{equation}
    \mathcal F = \frac{1}{1-K}\mathcal F_0,
    \label{F_(1-K)}
\end{equation}
where $\mathcal F_0$ is the zero-rung ladder, or a product of propagators.
To give this expression a concrete meaning, we diagonalize the kernel:

\begin{equation}
    K\circ \Psi_i=k_i \Psi_i.
    \label{Psi_def}
\end{equation}
Then in the basis of eigenfunctions of the kernel, the four-point function can be written as follows:

\begin{equation}
    \mathcal F = \sum_i \frac{1}{1-k_i} \frac{\langle \Psi_i, \mathcal F_0\rangle }{\langle \Psi_i, \Psi_i\rangle} \Psi_i.
    \label{F_sum}
\end{equation}

In the next sections, we proceed to find the eigenvalues $k_i$ and eigenfunctions $\Psi_i$ of the SYK kernel.
We find that these come in two sets distinguished by their symmetry under exchange of two fermions in (\ref{4pt_ij}).
The four-point function which is odd under $\tau_1 \leftrightarrow \tau_2$ and $\tau_3 \leftrightarrow \tau_4$ turns out to be the same as the four-point function of the SYK model with real fermions.
The full four-point function also contains a piece which is even under both of these symmetries.

\section{Discrete symmetries of the four-point function}
\label{sec:symmetries}

Our goal is to find the four-point function in the conformal limit of the SYK model:

\begin{equation}
    W\left( \tau_1,\tau_2,\tau_3,\tau_4 \right)\equiv\langle \bar{\psi}_i\left( \tau_1 \right) \psi_i\left( \tau_2 \right) \bar{\psi}_j\left( \tau_3 \right) \psi_j\left( \tau_4 \right) \rangle.
    \label{4pt_def}
\end{equation}

Following \cite{Maldacena:2016hyu} we expand this four-point function in a basis of the conformal Casimir.
But before doing that, let's look at discrete symmetries of $W$.

For the original SYK model with Majorana fermions, the four-point function is odd under exchange of the first two fermions:

\begin{equation}
    W^{\text{real}}\left( \tau_1,\tau_2,\tau_3,\tau_4 \right)=-W^{\text{real}}\left(\tau_2,\tau_1,\tau_3,\tau_4  \right).
    \label{W_real_sym}
\end{equation}
However for the model with complex fermions this is not so. 
If we write a complex fermion as a sum of two real ones:

\begin{equation}
    \psi=\xi+i\eta,
    \label{psi_real}
\end{equation}
then an arbitrary correlator containing two complex conjugated fermions looks like ($\left( \dots \right)$ standing for the terms independent of $\tau_1,\tau_2$):

\begin{multline}
  \left\langle \bar{\psi}\left( \tau_1 \right) \psi\left( \tau_2 \right)\left( \dots \right)\right\rangle = \\
    \left\langle \left(\xi \left( \tau_1 \right) \xi\left( \tau_2 \right) +\eta\left( \tau_1 \right)\eta\left( \tau_2 \right)\right)\left( \dots \right)\right\rangle+i\left\langle \left(\xi \left( \tau_1 \right) \eta\left( \tau_2 \right) -\eta\left( \tau_1 \right)\xi\left( \tau_2 \right)\right)\left( \dots \right)\right\rangle. 
    \label{corr_sum}
\end{multline}
The first term in the right-hand side is odd under $\tau_1 \leftrightarrow \tau_2$ and the second one is even:

\begin{equation}
    (1 \leftrightarrow 2)_{\text{even}}\equiv \xi \left( \tau_1 \right) \xi\left( \tau_2 \right) +\eta\left( \tau_1 \right)\eta\left( \tau_2 \right), \qquad (1 \leftrightarrow 2)_\text{odd}\equiv\xi \left( \tau_1 \right) \eta\left( \tau_2 \right) -\eta\left( \tau_1 \right)\xi\left( \tau_2 \right).
    \label{even_odd_def}
\end{equation}
So in particular the four-point function is a sum of two functions, one being odd and the other even under $\tau_1 \leftrightarrow \tau_2$. 
The same reasoning of course applies to another pair of fermions.

So naively, regarding discrete symmetries, the four-point function should look like:

\begin{multline}
    W\left( \tau_1,\tau_2,\tau_3,\tau_4 \right)=\langle(1 \leftrightarrow 2)_{\text{even}} \left( 3 \leftrightarrow 4 \right)_\text{even}\rangle -\langle(1 \leftrightarrow 2)_\text{odd} \left( 3 \leftrightarrow 4 \right)_\text{odd}\rangle +\\
    i\langle(1 \leftrightarrow 2)_\text{odd} \left( 3 \leftrightarrow 4 \right)_\text{even}\rangle +i\langle(1 \leftrightarrow 2)_\text{odd} \left( 3 \leftrightarrow 4 \right)_\text{even}\rangle.
    \label{odd_even}
\end{multline}
However, the two terms in the second line contain $i$, and $i$ is odd under the time-reversal symmetry $\mathcal T$.
The four-point function of a model with the Hamiltonian (\ref{H_def}) is $\mathcal T$-even in the large $N$ limit.
For $q=4k$, this is so because the Hamiltonian is manifestly $\mathcal T$-even.
For $q=4k+2$, the Hamiltonian contains $i$, but every ladder diagrams contains an even number of couplings, so the four-point function is again $\mathcal T$-even.

In the Appendix \ref{sec:app_T} we consider $\mathcal T$-odd four-point functions.
If we assume for now that the time-reversal symmetry is preserved, we can limit ourselves to considering only the first two terms in (\ref{odd_even}).
We give these terms superscripts $S$ and $A$, for the symmetric and antisymmetric parts:
\begin{equation}
    W\equiv W^S + W^A = \langle(1 \leftrightarrow 2)_{\text{even}} \left( 3 \leftrightarrow 4 \right)_\text{even}\rangle -\langle(1 \leftrightarrow 2)_\text{odd} \left( 3 \leftrightarrow 4 \right)_\text{odd}\rangle.
    \label{WSA_def}
\end{equation}
For the SYK model with Majorana fermions, the four-point function has only the antisymmetric part:
\begin{equation}
    W^{\text{real}}=W^A.
    \label{W_real_WA}
\end{equation}

The four-point function $W$ is conformally covariant.
To make it conformally invariant instead, we divide it by propagators:

\begin{equation}
    \mathcal W\left( \chi \right)\equiv \frac{W\left( \tau_1,\tau_2,\tau_3,\tau_4 \right)}{G\left( \tau_1,\tau_2 \right)G\left( \tau_3,\tau_4 \right)}=W^S\left( \chi \right)+ W^A\left( \chi \right).
    \label{W(chi)_def}
\end{equation}
Here $\mathcal W\left( \chi \right)$ is a function of the cross-ratio:
\begin{equation}
    \chi\equiv\frac{\tau_{12}\tau_{34}}{\tau_{13}\tau_{24}}.
    \label{chi_def}
\end{equation}
We can now fix the four coordinates in the standard way:

\begin{equation}
    \tau_1=0,\qquad \tau_2=\chi, \qquad \tau_3=1, \qquad \tau_4=\infty.
    \label{tau_fixed}
\end{equation}

In a conformal field theory invariant under time-reversal, the four-point function depends solely on the cross-ratio. 
However if we don't assume $\mathcal T$-invariance, the four-point function may also depend on the ordering of the points $\left( \tau_1, \tau_2, \tau_3,\tau_4\right)$.
The exchange of two points may reverse their cyclic ordering and therefore act as $\mathcal T$ on the four-point function.
Let's look at this more closely.

\begin{figure}
    \centering
    \psfig{file=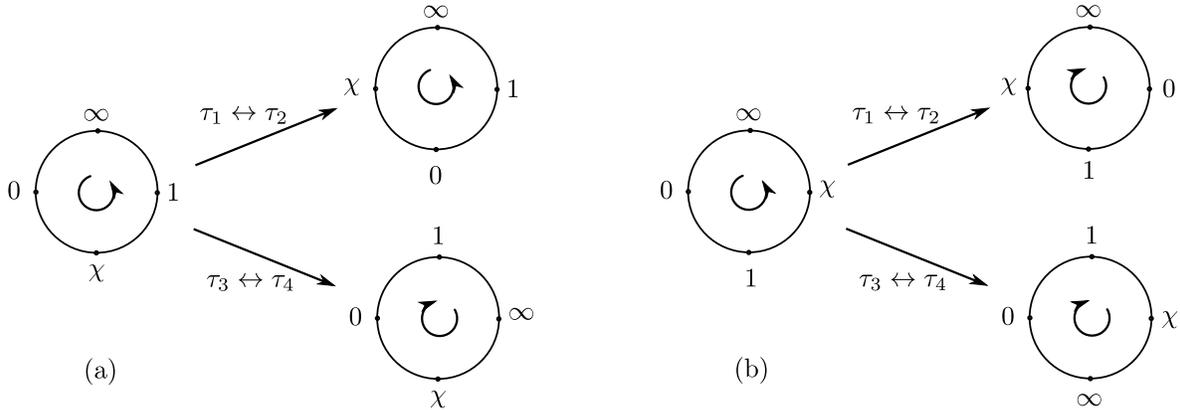, width=\textwidth}
    \caption{The action of exchange of coordinates on $\chi$ for (a) $0<\chi<1$ and (b) $\chi>1$. We can see that exchange of coordinates can reverse orientation together with taking $\chi \to \frac{\chi}{\chi-1}$.}
    \label{fig:inversion}
\end{figure}

What do discrete symmetries imply for $\mathcal W\left( \chi \right)$?
Naively, both the exchanges $\tau_1 \leftrightarrow \tau_2$ and $\tau_3 \leftrightarrow \tau_4$ act as:
\begin{equation}
    \chi \to \frac{\chi}{\chi-1}.
    \label{chi_to}
\end{equation}
However, these transformations can reverse the orientation of time, depending on the value of $\chi$. 
From the fig. \ref{fig:inversion} we see that the exchanges of coordinates act as:

\begin{equation}
    \begin{matrix}
        &\chi<1 & \,&\chi>1\\
        \tau_1 \leftrightarrow \tau_2 & \chi \to \frac{\chi}{\chi-1} & \,&\left( \chi \to \frac{\chi}{\chi-1} \right) \circ \mathcal T\\
        \tau_3 \leftrightarrow \tau_4 & \left( \chi \to \frac{\chi}{\chi-1} \right) \circ \mathcal T&\,&\left( \chi \to \frac{\chi}{\chi-1} \right) \circ \mathcal T\\
    \end{matrix}
    \label{inversions}
\end{equation}

This means that for a $\mathcal T$-invariant theory, the exchange of coordinates acts exactly like $\chi \to \frac{\chi}{\chi-1}$, and therefore:

\begin{equation}
    \mathcal W^A\left( \frac{\chi}{\chi-1} \right)=\mathcal W^A\left( \chi \right), \qquad \mathcal W^S\left( \frac{\chi}{\chi-1} \right)=-\mathcal W^S\left( \chi \right).
    \label{W_chi_to}
\end{equation}
Note that under the transformation (\ref{chi_to}), the antisymmetric part is even and the symmetric part is odd, because the propagators (\ref{G_def}) are odd under the exchange of fermions.

The $\mathcal T$-odd four-point functions should transform oppositely under exchanges $\tau_1 \leftrightarrow \tau_2$ and $\tau_3 \leftrightarrow \tau_4$. 
For $\chi<1$, these transformations acts oppositely on a $\mathcal T$-odd function.
There are two options: the four-point function can be either odd or even under $\chi \to \frac{\chi}{\chi-1}$, depending on whether it is even or odd under $\tau_1 \leftrightarrow \tau_2$.
For $\chi>1$ however, these exchanges act in the same way, so the $\mathcal T$-odd four-point function has to be zero in that region.
In Appendix \ref{sec:app_T} we will see this from a direct calculation.

%

In the next section, we define the SYK kernel for the model with complex fermions and find its eigenfunctions using the fact that it commutes with the Casimir of the conformal group.
The kernel is therefore diagonalized by three-point functions, which can also be odd or even under these discrete symmetries.
Then using the shadow formalism, we construct the four-point functions out of the three-point functions with the desired symmetries.

\section{Eigenvalues of the kernel}
\label{sec:kernel}

\begin{figure}
    \centering
    \psfig{file=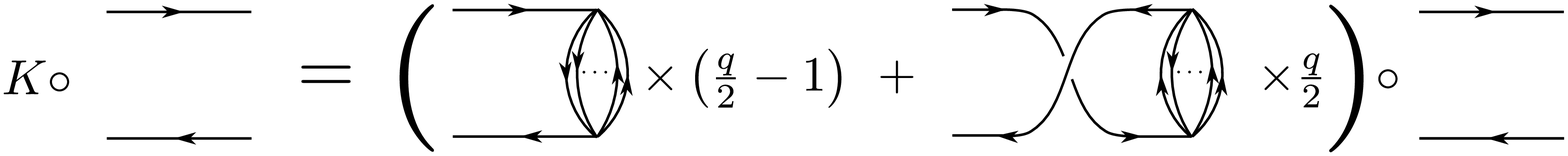, width=\textwidth}
    \caption{The kernel for complex fermions. The first term has $\frac{q}{2}-1$ arrows pointing up and $\frac{q}{2}-1$ arrows pointing down in the rung; the second term has $\frac{q}{2}$ arrows pointing up and only $\frac{q}{2}-2$ arrows pointing down.}
    \label{fig:complex_kernel}
\end{figure}

The SYK kernel is an integral operator acting on the four-point functions, corresponding to adding one rung to a ladder diagram (fig. \ref{fig:kernel}).
For the complex model, one can represent the kernel schematically as in fig. \ref{fig:complex_kernel}.
There are two types of rungs which can be added, one coming with a factor of $\left( \frac{q}{2}-1 \right)$ and the other with a factor of $\frac{q}{2}$, reflecting the choice of an index going down the ``rail'' of the ladder.

The kernel commutes with the action of the conformal algebra.
Specifically, if we define the $sl_2$ generators as:

\begin{eqnarray}
    L_0^{(\tau)}&=& -\tau\partial_\tau - \Delta, \\
    L_{-1}^{(\tau)}&=& -\partial_\tau,\\
    L_1^{(\tau)}&=& -\tau^2 \partial_\tau - 2\Delta \tau,
    \label{L_i_def}
\end{eqnarray}
the kernel obeys the following:

\begin{equation}
    \left( L_i^{(1)}+L_i^{(2)} \right)K\left( \tau_1,\tau_2; \tau_3,\tau_4 \right)=K\left( \tau_1,\tau_2;\tau_3,\tau_4 \right)\left( L_i^{(3)}+L_i^{(4)} \right), \qquad i=1,2,3.
    \label{sl2_kernel}
\end{equation}
In particular, this means that the kernel commutes with the two-particle Casimir:
\begin{equation}
    C^{\left( 12 \right)}\equiv\left( L_0^{(1)}+L_0^{(2)} \right)^2-\frac12 \left\{ L_{-1}^{(1)}+L_{-1}^{(2)},  L_{1}^{(1)}+L_{1}^{(2)} \right\},
    \label{C_1+2_def}
\end{equation}
and therefore, the kernel and the Casimir have a common basis of eigenfunctions.

The Casimir is diagonalized by conformal three-point functions. 
For our purposes, we are most interested in the three-point functions of two complex conjugated fermions and a bosonic operator of dimension $h$:

\begin{equation}
    \left\langle \bar{\psi}\left( \tau_1 \right) \psi\left( \tau_2 \right) V_h\left( \tau_0 \right)\right\rangle.
    \label{3pt_def}
\end{equation}
Depending on the operator $V_h$, this three-point function may be symmetric or antisymmetric in $\left( \tau_1,\tau_2 \right)$.
For example, if $V_h$ is the identity, we have:
\begin{equation}
    \left\langle \bar{\psi}\left( \tau_1 \right) \psi\left( \tau_2 \right) 1\right\rangle=\frac{\sgn \left( \tau_1-\tau_2 \right)}{\left|\tau_1-\tau_2\right|^{2\Delta}},
    \label{3pt_1}
\end{equation}
which is antisymmetric under exchange of fermions, while for $V_h=\bar{\psi}\psi$:
\begin{equation}
    \left\langle \bar{\psi}\left( \tau_1 \right) \psi\left( \tau_2 \right) \bar{\psi}\psi\left( \tau_0 \right)\right\rangle=\frac{\sgn \left( \tau_1-\tau_0 \right) \sgn \left( \tau_2-\tau_0 \right)}{\left|\tau_1-\tau_2\right|^{2\Delta}},
    \label{3pt_charge}
\end{equation}
which is symmetric under the same exchange. 
(Here we have used the fact that $\bar{\psi}\psi$ is a conserved charge and hence it has dimension zero.)
A generic three-point function is a sum of an antisymmetric and a symmetric part, which we call $f^A_h$ and $f^S_h$.
With a suitable normalization of $V_h$, the three-point function looks like:

\begin{equation}
    \left\langle \bar{\psi}\left( \tau_1 \right) \psi\left( \tau_2 \right) V_h\left( \tau_0 \right)\right\rangle=f^A_h+if^S_h=\frac{\sgn \left( \tau_1-\tau_2 \right) + i \sgn \left( \tau_1-\tau_0 \right) \sgn \left( \tau_2-\tau_0 \right)}{\left|\tau_1-\tau_2\right|^{2\Delta - h}\left|\tau_1-\tau_0\right|^{h}\left|\tau_1-\tau_0\right|^{h}}.
    \label{3pt_formula}
\end{equation}
This three-point function is an eigenfunction of the Casimir with eigenvalue $h\left( h-1 \right)$:

\begin{equation}
    C^{(12)}\left\langle \bar{\psi}\left( \tau_1 \right) \psi\left( \tau_2 \right) V_h\left( \tau_0 \right)\right\rangle=h\left( h-1 \right)\left\langle \bar{\psi}\left( \tau_1 \right) \psi\left( \tau_2 \right) V_h\left( \tau_0 \right)\right\rangle,
    \label{C_3pt}
\end{equation}
and therefore it is an eigenfunction of the kernel.
For simplicity let's define separately the kernels acting on the symmetric and antisymmetric three-point functions. 
They differ by a factor of $(q-1)$:

\begin{equation}
    K^A\left( \tau_1,\tau_2;\tau_1',\tau_2' \right)=\left( q-1 \right)K^S=-J^2\left( q-1 \right) G\left( \tau_1,\tau_1' \right)G\left( \tau_2,\tau_2' \right)G^{q-2}\left( \tau_1',\tau_2' \right)d\tau_1'd\tau_2',
    \label{KA_def}
\end{equation}

We want to find eigenvalues of $K^A$:
\begin{equation}
    K^A \circ f^A_h=\int K^A f^A_h=k^A(h) f^A_h.
    \label{KA_fh}
\end{equation}
Since we already know that the three-point functions $f^A_h$ diagonalize the kernel, we can consider a convenient limit of this expression.
Taking the position of the boson to infinity,

\begin{equation}
    f^A_h\left( \tau_1,\tau_2,\tau_0 \right) \xrightarrow[\tau_0 \to\infty ]{} \tau_0^{-2h} \frac{\sgn \tau_{12}}{|\tau_{12}|^{2\Delta-h}}, 
    \label{fA_infty}
\end{equation}
and fixing the coordinates in the kernel to be 0 and 1, we can compute the eigenvalue as follows:

\begin{equation}
    k^A\left( h \right)= \left.\tau_0^{2h}  \int d\tau_1' d\tau_2' K \left( 1,0;\tau_1',\tau_2'  \right) f^A_h\left( \tau_1',\tau_2',\tau_0 \right)\right|_{\tau_0 \to \infty}.
    \label{k_c_comp}
\end{equation}
Using the explicit form of the kernel and changing variables (see Appendix \ref{sec:app_kernel}), we recast this integral in a symmetric form:
\begin{equation}
    k^A\left( h \right)=\frac{1}{\alpha_0} \int d\tau \frac{\sgn \tau}{|\tau|^{2\Delta} |1-\tau|^{1-h}}\int d\tau' \frac{\sgn \tau'}{|\tau'|^{2\Delta}|1-\tau'|^{h}},
    \label{k_c_comp_sym}
\end{equation}
where
\begin{equation}
    \frac1{\alpha_0}={\left( q-1 \right)J^2 b^q}=\left( 1-\Delta \right)\left( 1-2\Delta \right)\frac{\tan \pi \Delta}{2\pi \Delta}.
    \label{alpha_def}
\end{equation}

Taking this integral, we arrive at a relatively simple expression in terms of Gamma-functions:

\begin{equation}
    k^A\left( h, \Delta\right)=\frac{1}{\pi} \frac{\Gamma\left( -2\Delta \right)}{\Gamma\left( 2\Delta -2\right)}\Gamma\left( 2\Delta-h \right)\Gamma\left( 2\Delta+h-1 \right)\left(\sin \pi h - \sin 2\pi \Delta \right).
    \label{kA_ans}
\end{equation}
This  eigenvalue has been found in a slightly different form in \cite{Maldacena:2016hyu}.

In \cite{Maldacena:2016hyu}, it was argued that the theory is exactly solvable for large $q$ and for $q=2$.
In these limits, the eigenvalue of the kernel was found to be:

\begin{eqnarray}
    q=\infty:& \qquad &k^A\left( h, 0 \right)=\frac{2}{h(h-1)},\\
    q=2: &\qquad &  k^A\left( h,\frac{1}{2} \right)=-1.
    \label{kA_limits}
\end{eqnarray}

For a minimal ``generic'' case of $q=4$, 
\begin{equation}
    q=4: \qquad k^A\left( h,\frac{1}{4} \right)=-\frac{3}{2}\frac{\tan \frac{\pi}{2} \left( h-\frac12 \right)}{h-\frac12}.
    \label{kA_Delta14}
\end{equation}

To find the dimensions of the operators in the model, we have to solve the condition:

\begin{equation}
    k^A(h,\Delta)=1.
    \label{kA=1}
\end{equation}
The eigenvalue (\ref{kA_ans}) is symmetric under $h \leftrightarrow 1-h$.
The dimensions of the physical operators in the spectrum are positive, so we keep only one copy with positive $h$.
This is justified by a procedure of changing the integration contour in \cite{Maldacena:2016hyu}.
For $q>2$, the eigenvalue equation has infinitely many seemingly irrational solutions, given by an asymptotic formula:
\begin{equation}
  h=2k+1+2\Delta+ O\left( \frac{1}{k} \right),\qquad k>0.
    \label{h_A}
\end{equation}
which allows us to identify them with the operators:
\begin{equation}
    \mathcal O_k^A=\bar{\psi} \partial^{2k+1}_\tau\psi, \qquad k\ge 0.
    \label{OA_def}
\end{equation}
There is also an integer solution at $h=2$, by (\ref{OA_def}) corresponding to the operator:

\begin{equation}
    \mathcal O^A_{h=2}=\bar \psi \partial_\tau \psi.
    \label{OA_1_def}
\end{equation}
This operator can be understood as a (pseudo)Goldstone boson for the broken reparameterization symmetry. 

Following the same steps for the symmetric three-point function, we arrive at a very similar integral expression for the eigenvalue of the kernel:

\begin{equation}
    k^S\left( h \right)=\frac{1}{\alpha_0}\frac{1}{q-1} \int d\tau \frac{\sgn \tau \sgn \left( 1-\tau \right)}{|\tau|^{2\Delta} |1-\tau|^{1-h}}\int d\tau' \frac{\sgn \tau' \sgn\left( 1-\tau' \right)}{|\tau'|^{2\Delta}|1-\tau'|^{h}},
    \label{kB_int}
\end{equation}
which gives the following answer:
\begin{equation}
    k^S\left( h,\Delta \right)=\frac{1}{\pi}\frac{\Gamma\left(1 -2\Delta \right)}{\Gamma\left( 2\Delta-1 \right)}\Gamma\left( 2\Delta-h \right)\Gamma\left( 2\Delta+h-1 \right)\left( \sin \pi h + \sin 2 \pi \Delta \right).
    \label{kB_ans}
\end{equation}

At the same limiting solvable cases, this eigenvalue is:

\begin{eqnarray}
  q \to \infty: & \qquad &k^S\left( h,0 \right)=\frac{2}{qh(h-1)}+O\left( q^{-2} \right),\\
    q \to 2: & \qquad & k^S \left( h,\frac{1}{2} \right)= -1.
    \label{kB_limits}
\end{eqnarray}
We notice that for large $q$, the symmetric and antisymmetric kernels differ by a factor of $q$.
This fact will allow us to find the correction to the eigenvalue of the $h=1$ mode in Section \ref{sec:h=1}.

The case of $q=4$ has been addressed in \cite{Klebanov:2016xxf}, \cite{Gross:2016kjj}.
\begin{equation}
    k^S\left( h, \frac{1}{4} \right)= \frac{1}{2} \frac{\cot \frac{\pi}{2} \left( h-\frac12 \right)}{h-\frac12}.
    \label{kB_4}
\end{equation}

Solving the condition:
\begin{equation}
    k^S(h,\Delta)=1,
    \label{kB=1}
\end{equation}
again gives an infinite set of operators with (positive) irrational dimensions:
\begin{equation}
    h=2k+2\Delta+O\left( \frac{1}{k} \right),
    \label{h_S}
\end{equation}
which we can identify as:

\begin{equation}
    \mathcal O^S_k=\bar{\psi} \partial^{2k}_\tau \psi, \qquad k\ge 0.
    \label{OB_def}
\end{equation}
One operator in this series has an integer dimension $h=0$, for all $q>2$.
It is the $U(1)$ charge:
\begin{equation}
  O^S=\bar{\psi}\psi,
    \label{O_h=0}
\end{equation}
As expected, the conserved charge is not renormalized and has vanishing dimension in the near-conformal limit as well.
A similar procedure of changing the integration contour allows us to divide the spectrum by the $h \leftrightarrow 1-h$ symmetry and keep only positive dimensions of the operators.
In that case, the $U(1)$ charge gets identified with the $h=1$ mode.

In the Section \ref{sec:chaos}, we will find that the three-point functions with these two special operators with integer dimensions, $\bar{\psi} \partial_\tau \psi$ and $\bar{\psi}\psi$, are eigenfunctions of the retarded kernel and therefore can contribute to the chaotic behavior.
But while the operator with $h=2$ has a maximal Lyapunov exponent in the sense of the bound of \cite{Maldacena:2015waa}, the $U(1)$ charge has a zero Lyapunov exponent.

In the next section, we find the basis for the conformal four-point functions in the shadow representation.

\section{Four-point function in the shadow formalism}
\label{sec:shadow}

As we have seen above, the leading in $1/N$ correction to the four-point function comes from ladder diagrams.
We define a conformally invariant version of this correction, dividing by propagators:

\begin{equation}
    \mathcal F\left( \chi \right)=\frac{\mathcal F \left( \tau_1,\tau_2,\tau_3,\tau_4 \right)}{G\left( \tau_1,\tau_2 \right)G\left( \tau_3,\tau_4 \right)}.
    \label{F(chi)_def}
\end{equation}
This function can be expanded in the basis of eigenfunctions of the two-particle Casimir.
The Casimir $C^{\left( 12 \right)}$ can be rewritten in terms of the cross-ratio:

\begin{equation}
    \mathcal C \left( \chi \right)\equiv \chi^2 \left( 1-\chi \right)\partial_\chi^2 - \chi^2 \partial_\chi, \qquad C^{\left( 12 \right)} \mathcal F\left( \tau_1,\tau_2,\tau_3,\tau_4 \right)=\mathcal C\left( \chi \right) \mathcal F\left( \chi \right).
    \label{C_def}
\end{equation}

Eigenfunctions of $\mathcal C$ are the $sl_2$ conformal blocks, which we call $F_h(\chi)$ (having chosen a convenient normalization):

\begin{equation}
    F_h\left( \chi \right)\equiv \frac{\Gamma^2(h)}{\Gamma(2h)} \chi^h \,_2F_1\left( h,h;2h;\chi \right), \qquad \chi<1.
    \label{Fh_def}
\end{equation}
The eigenvalues of the Casimir are of course the same $h(h-1)$ we have seen when discussing three-point functions:
\begin{equation}
    \mathcal C(\chi)F_h=h(h-1)F_h.
    \label{CF_h}
\end{equation}
Note that the eigenvalue is symmetric under $h \leftrightarrow 1-h$.
Given this symmetry and the fact that the Casimir is a second-order differential operator, its generic eigenfunction is a linear combination of $F_h, F_{1-h}$:
\begin{equation}
    \Psi_h=a(h) F_h(\chi)+ a(1-h) F_{1-h}\left( \chi \right).
    \label{aF_h}
\end{equation}
For a given $h$, we can adjust $a(h)$ to make this combination odd or even under $\chi \to \frac{\chi}{\chi-1}$.
These two options span respectively symmetric and antisymmetric four-point functions. 
This means that having found the contributions to the four-point functions of both types for each $h$, we cover all the eigenspace of the conformal Casimir. 

To make the Casimir a Hermitean operator, we have to make sure its eigenvalues are real.
This leaves us with two choices:
\begin{equation}
    h=\frac12 + is, \qquad \text{or} \qquad h \in \mathbb R.
    \label{h_is}
\end{equation}
In fact, not all real values of $h$ are allowed.
This is because for real $h$ the eigenfunction $F_h$ has a monodromy around zero.
Since we are also interested in symmetry under $\chi \to \frac{\chi}{\chi-1}$ to fix the discrete symmetries of the four-point function, we write it as a ``half-monodromy'':

\begin{equation}
    F_h \left( \frac{\chi}{\chi-1} \right)=e^{\pi i h} F_h\left( \chi \right), \qquad 0<\chi<1.
    \label{Fh_monodromy}
\end{equation}
From here we see two things. 
First, the continuous series $h \in \mathbb R$ is reduced to a discrete series of ``bound states'':

\begin{equation}
    h \in \mathbb Z.
    \label{h_Z}
\end{equation}
Second, in this integer series the antisymmetric four-point function (in the sense of \ref{W_chi_to}) should have even $h$:
\begin{equation}
    \Psi_h^A = a^A(h) F_h\left( \chi \right), \qquad h \in 2 \mathbb Z,
    \label{h_int_A}
\end{equation}
and the symmetric four-point function should have odd $h$:
\begin{equation}
    \Psi_h^S = a^S(h) F_h\left( \chi \right), \qquad h \in 2 \mathbb Z+1,
    \label{h_int_S}
\end{equation}
We will confirm this later while considering normalization conditions on $\Psi_h^A, \Psi_h^S$.

The antisymmetric eigenfunction $\Psi^A_h$ has been found in \cite{Maldacena:2016hyu}. 
Our next step is to find the explicit expression for the symmetric one $\Psi^S_h$.

\subsection{Shadow formalism}

We have seen that the conformal Casimir is diagonalized by three-point functions of the form $\left \langle \bar{\psi}\psi V_h\right\rangle$, and it is also diagonalized by the four-point function $\Psi_h\left( \chi \right)$ (up to a product of propagators).
So it seems natural to suggest that the conformal four-point function $\Psi_h\left( \chi \right)$ is constructed out of three-point functions.
This idea is embodied in the shadow formalism \cite{ferrara1972shadow},\cite{SimmonsDuffin:2012uy}. 
It works as follows.

Consider the four-point function to be consisting of two parts which belong to two different decoupled CFT:
\begin{equation}
    \langle \bar{\psi}\left( t_1 \right) \psi\left( t_2 \right) \bar{\psi}\left( t_3 \right) \psi\left( t_4 \right) \rangle\to \langle \bar{\psi}\left( t_1 \right) \psi\left( t_2 \right)\rangle_1 \langle \bar{\psi}\left( t_3 \right) \psi\left( t_4 \right)\rangle_2.
    \label{4pt_trivial}
\end{equation}
Now let's add a small interaction.
Let's introduce an operator $V$ living in the first CFT and an operator $V'$ living in the second CFT, coupled slightly via:
\begin{equation}
    S \supset \varepsilon \int dt_0 V_h(t_0) V'_{1-h}(t_0). 
    \label{lagr_pert}
\end{equation}
To make this interaction conformal, we fix the sum of dimensions of these two operators to be one.
Then to the first order in $\varepsilon$ the four-point function is:
\begin{multline}
    \langle \bar{\psi}\left( t_1 \right) \psi\left( t_2 \right) \bar{\psi}\left( t_3 \right) \psi\left( t_4 \right) \rangle=
    \langle \bar{\psi}\left( t_1 \right) \psi\left( t_2 \right)\rangle\langle \bar{\psi}\left( t_3 \right) \psi\left( t_4 \right) \rangle+\\
    \varepsilon \sum_h\int dt_0 \langle \bar{\psi}\left( t_1 \right) \psi\left( t_2 \right) V_h\left( t_0 \right)\rangle \langle \bar{\psi}\left( t_3 \right) \psi\left( t_4 \right) V'_{1-h}\left( t_0\right)\rangle.
    \label{4pt_shadow}
\end{multline}
Comparing this to (\ref{F4pt_def}) and (\ref{F(chi)_def}), we find an integral expression for the eigenfunctions of the Casimir:

\begin{equation}
    \Psi_h\left( \chi \right) =\frac{1}{G\left( \tau_1,\tau_2 \right) G\left( \tau_3,\tau_4 \right)}\int d\tau_0 {\langle \bar{\psi}\left( \tau_1 \right) \psi\left( \tau_2 \right) V_h\left( \tau_0 \right)\rangle \langle \bar{\psi}\left( \tau_3 \right) \psi\left( \tau_4 \right) V'_{1-h}\left( \tau_0\right)\rangle}.
    \label{F_shadow}
\end{equation}
The four-point function inherits the discrete symmetries of the three-point functions.
This allows us to readily write the expressions for $\mathcal T$-even symmetric and antisymmetric four-point functions:

\begin{eqnarray}
    \label{PsiA_ff}
    \Psi^{A}_h\left( \chi \right)&=& \int d\tau_0 \frac{f^A_h \left( \tau_1,\tau_2,\tau_0 \right) f^A_{1-h}\left( \tau_3,\tau_4,\tau_0 \right)}{G\left( \tau_1,\tau_2 \right)G\left( \tau_3,\tau_4 \right)},\\
    \label{PsiB_ff}
    \Psi^{S}_h\left( \chi \right)&=& \int d\tau_0 \frac{f^S_h \left( \tau_1,\tau_2,\tau_0 \right) f^S_{1-h}\left( \tau_3,\tau_4,\tau_0 \right)}{G\left( \tau_1,\tau_2 \right)G\left( \tau_3,\tau_4 \right)},    
\end{eqnarray}
with the three-point functions $f^S_h, f^A_h$ defined in (\ref{3pt_formula}).
These expressions are manifestly symmetric under $h \leftrightarrow 1-h$, which is to be expected since the eigenvalues of the Casimir are symmetric under this too.

The antisymmetric and symmetric eigenfunctions are respectively even and odd under $\chi \to \frac{\chi}{\chi-1}$:
\begin{equation}
    \Psi^A_h\left( \frac{\chi}{\chi-1} \right)=\Psi^A_h\left( \chi \right), \qquad \Psi^S_h\left( \frac{\chi}{\chi-1} \right)=-\Psi^S_h\left( \chi \right).
    \label{Psi_chi_to}
\end{equation}

Taking the two integrals (\ref{PsiA_ff}, \ref{PsiB_ff}), we find the explicit form of the eigenfunctions.
To do this conveniently, let's introduce one more function:

\begin{equation}
    G_h\left(\frac1{\chi} \right)=\frac{2\pi}{\sin \pi h} \,_2F_1\left( h,1-h;1; \frac1{\chi}\right), \qquad \chi<0 \qquad \text{or} \qquad \chi>1.
    \label{Gh_def}
\end{equation}
This function is proportional to the difference $\left( F_h-F_{1-h} \right)$ in the regions where both are defined:

\begin{eqnarray}
    G_h\left( \frac{1}{\chi} \right)&=& \frac1{2\pi}\tan \pi h\left( F_h\left( \frac{\chi}{\chi-1} \right)-F_{1-h}\left( \frac{\chi}{\chi-1} \right) \right), \qquad \chi<0\\
    G_h\left( 1-\frac{1}{\chi} \right)&=& \frac1{2\pi}\tan \pi h\left( F_h\left( \chi \right)-F_{1-h}\left( \chi \right) \right), \qquad 0<\chi<1.\\
    \label{Gh_Fh}
\end{eqnarray}

%
%
%

%

Unlike the $F_h$ defined in (\ref{Fh_def}), this function is symmetric under $h \leftrightarrow 1-h$.

The even four-point function is the same as the four-point function of the real SYK model \cite{Maldacena:2016hyu}:
\begin{equation}
    \Psi^{A}_h\left(\chi  \right)=\left\{
    \begin{gathered}
        F_h\left( \chi \right)+F_{1-h}\left( \chi \right)+G_h\left( \frac{1}{\chi} \right), \qquad \chi<0,\\
        F_h\left( \chi \right)+F_{1-h}\left( \chi \right)+G_h\left( \frac{\chi-1}{\chi} \right), \qquad 0<\chi<1,\\
        G_h\left( \frac{\chi-1}{\chi} \right)+G_h\left( \frac{1}{\chi} \right), \qquad \chi>1.
    \end{gathered}
    \right.
    \label{IAA_FG}
\end{equation}
The odd function is as follows:

\begin{equation}
    \Psi^{S}_h\left(\chi  \right)=\left\{
    \begin{gathered}
        F_h\left( \chi \right)+F_{1-h}\left( \chi \right)-G_h\left( \frac{1}{\chi} \right), \qquad \chi<0,\\
        -F_h\left( \chi \right)-F_{1-h}\left( \chi \right)+G_h\left( \frac{\chi-1}{\chi} \right), \qquad 0<\chi<1,\\
        G_h\left( \frac{\chi-1}{\chi} \right)-G_h\left( \frac{1}{\chi} \right), \qquad \chi>1,
    \end{gathered}
    \right.
    \label{IBB_FG}
\end{equation}
This form makes it clear that both these integrals are symmetric under $h \leftrightarrow 1-h$ as the shadow integral (\ref{4pt_shadow}) suggests, and also that $\Psi^{S}$ is symmetric under $\chi \to \frac{\chi}{\chi-1}$ and $\Psi^{A}$ is anti-symmetric under that.

Making use of various hypergeometric identities, we can recast these expressions as:

\begin{equation}
    \Psi^{A}_h\left(\chi  \right)=\left\{
    \begin{gathered}
        \frac{2}{\cos \pi h}\left(  \cos^2 \frac{\pi h}{2}F_h(\chi) -  \sin^2 \frac{\pi h}{2} F_{1-h}(\chi)\right), \qquad \chi<1,\\
        \frac{2}{\sqrt{\pi}} \Gamma\left( \frac{h}{2} \right) \Gamma\left( \frac{1-h}{2} \right)\,_2F_1\left( \frac{h}{2},\frac{1-h}{2};\frac{1}{2}; \frac{\left( 2-\chi \right)^2}{\chi^2} \right), \qquad \chi>1.
    \end{gathered}
    \right.
    \label{IAA_2F1}
\end{equation}
and:

\begin{equation}
    \Psi^{S}_h\left(\chi  \right)=\left\{
        \begin{aligned}[r]
        \frac{2}{\cos \pi h}\left(    \sin^2 \frac{\pi h}{2}F_h(\chi)- \cos^2 \frac{\pi h}{2} F_{1-h}(\chi)\right), \qquad \chi<1,\\
        \frac{4}{\sqrt{\pi}} \left( \frac{2-\chi}{\chi} \right)\Gamma\left( 1-\frac{h}{2} \right)\Gamma\left( \frac{1+h}{2} \right) \,_2F_1\left( 1-\frac{h}{2},\frac{1+h}{2};\frac{3}{2}; \frac{\left( 2-\chi \right)^2}{\chi^2} \right), \\
        \chi>1.
    \end{aligned}
    \right.
    \label{IBB_2F1}
\end{equation}
%
\subsection{Normalization and bound states}
\label{sec:norm}

To express the four-point function in terms of eigenfunctions of the Casimir (\ref{IAA_2F1}, \ref{IBB_2F1}), we have to find inner products between eigenfunctions. 
We define the inner product as:
\begin{equation}
    \langle g,f \rangle = \int_{-\infty}^\infty \frac{d\chi}{\chi^2}\bar{g}(\chi) f(\chi).
    \label{norm_def}
\end{equation}
The Casimir operator is Hermitean with respect to this product (up to boundary terms).
The eigenvalues of the Casimir $h(h-1)$ ought to be real, hence we expect $h$ either to be integer or to belong to the continuous series $h=\frac12+is$.
It is easy to see that for non-integer real $h$, the norm defined in (\ref{norm_def}) diverges.

For the continuous series, the inner product is:

\begin{equation}
    \langle \Psi^{A}_h, \Psi^{A}_{h'}\rangle=\langle \Psi^{S}_h, \Psi^{S}_{h'}\rangle=\frac{\pi \coth \pi s}{s}4\pi \delta\left( s-s' \right).
    \label{s_prod}
\end{equation}
In this case, the main contribution to the integral (\ref{norm_def}) comes from the region $\chi \sim 0$. 
In this region, the four-point function looks like:
\begin{equation}
    \Psi^A_h\left( \chi \right) \sim \left( 1+\frac{1}{\cos \pi h} \right)\frac{\Gamma^2\left( h \right)}{\Gamma\left( 2h \right)}\chi^h+\left( h \leftrightarrow 1-h \right).
    \label{psi_chi_0}
\end{equation}
Using the integral form of delta-function:
\begin{equation}
    \int_{-\infty}^\infty\frac{d\chi}{\chi}\left( \chi^{i\left( s-s' \right)}+\chi^{-i\left( s-s' \right)} \right)=4\pi \delta\left( s-s' \right),
    \label{delta_int}
\end{equation}
we find that the inner product is:
\begin{equation}
    \langle \Psi^{A}_h, \Psi^{A}_{h'}\rangle=4\pi \delta\left( s-s' \right)\left( 1+\frac{1}{\cos \pi h} \right)\frac{\Gamma^2\left( h \right)}{\Gamma\left( 2h \right)}\cdot \left( h \leftrightarrow 1-h \right)=\frac{\pi \coth \pi s}{s}4\pi \delta\left( s-s' \right),
    \label{psi_psi_exp}
\end{equation}
with the same expression for inner product of symmetric states.

The states with different symmetry are of course orthogonal:
\begin{equation}
    \langle \Psi^{A}_h, \Psi^{S}_{h'}\rangle=0.
    \label{s_ortho}
\end{equation}

For integer values of $h$, there are two options, either $h$ is even positive, or odd negative:
\begin{equation}
    h\in 2\mathbb Z,\qquad h\ge 2 \qquad \text{or} \qquad h\in 1+2\mathbb Z,\qquad  h\le -1.
    \label{h_even_def}
\end{equation}
For this series, the antisymmetric eigenfunction is normalizable,

\begin{equation}
    \langle \Psi^{A}_{h'}, \Psi^{A}_{h}\rangle = \frac{2\pi^2 \delta_{hh'}}{\left|h-\frac12\right|},
    \label{Ih_even}
\end{equation}
while the symmetric function diverges, as can easily be seen from its form at $\chi>1$.
If we divide the spectrum by the $h \leftrightarrow 1-h$ symmetry, there are only even positive states left, which agrees with the expectation from the half-monodromy (\ref{h_int_A}):
\begin{equation}
  h^A \in 2\mathbb Z_+.
  \label{h_A_even}
\end{equation}

The second series is the complement of the first one:
\begin{equation}
    h \in1+2 \mathbb Z, \qquad h \ge 1 \qquad \text{or} \qquad h\in 2\mathbb Z, \qquad h \le 0.
    \label{h_odd_def}
\end{equation}
In this case, the symmetric function is normalizable, and the antisymmetric is not:

\begin{equation}
    \langle \Psi^{S}_{h'}, \Psi^{S}_{h}\rangle = \frac{2\pi^2 \delta_{hh'}}{\left|h-\frac12\right|}.
    \label{Ih_odd}
\end{equation}
Here we also can get rid of the non-positive dimensions in the spectrum, leaving only:
\begin{equation}
  h^S \in 2\mathbb Z_++1,
  \label{h_S_odd}
\end{equation}
in agreement with the half-monodromy argument (\ref{h_int_S}).

\begin{figure}
    \centering
    \psfig{file=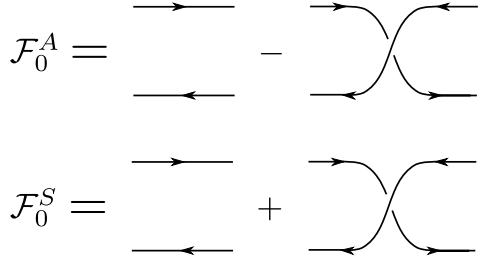, width=.4\textwidth}
    \caption{Odd and even zero-rung four-point functions.}
    \label{fig:0rung}
\end{figure}

Now we can express the four-point function as a sum using the formula \ref{F_sum}.
In order to do that, we need an expression for the zero-rung four-point function.
The zero-rung four-point function is a product of propagators, but it can be even or odd under exchange of fermions (see fig. \ref{fig:0rung}):

\begin{eqnarray}
    \mathcal F_0^{A}\left(\tau_1,\tau_2,\tau_3, \tau_4\right)&=&G\left( \tau_1,\tau_3 \right) G\left(\tau_2,\tau_4 \right)-G\left( \tau_1,\tau_4 \right)G\left( \tau_2,\tau_3 \right),\\
    \mathcal F_0^{S}\left(\tau_1,\tau_2,\tau_3, \tau_4\right)&=&G\left( \tau_1,\tau_3 \right) G\left(\tau_2,\tau_4 \right)+G\left( \tau_1,\tau_4 \right)G\left( \tau_2,\tau_3 \right).
    \label{F0_def}
\end{eqnarray}
Making them conformally invariant, we find an expression in terms of cross-ratio $\chi$:

\begin{eqnarray}
   \Psi_0^{A}\left( \chi \right)&=& \frac{\mathcal F_0^A \left( \tau_1,\tau_2,\tau_3,\tau_4 \right)}{G\left( \tau_1,\tau_2 \right)G\left( \tau_3,\tau_4 \right)}=-\sgn \chi \cdot|\chi|^{2\Delta}+\sgn \chi \sgn \left( 1-\chi \right)\left|\frac{\chi}{\chi-1}\right|^{2\Delta},\\
  \Psi_0^{S}\left( \chi \right)&=& \frac{\mathcal F_0^S \left( \tau_1,\tau_2,\tau_3,\tau_4 \right)}{G\left( \tau_1,\tau_2 \right)G\left( \tau_3,\tau_4 \right)}=-\sgn \chi \cdot|\chi|^{2\Delta}-\sgn \chi \sgn \left( 1-\chi \right)\left|\frac{\chi}{\chi-1}\right|^{2\Delta}.
    \label{Psi0_def}
\end{eqnarray}

The $\Psi_0^A$, $\Psi_0^S$ zero-rung four-point functions are respectively even and odd under $\chi \to \frac{\chi}{\chi-1}$, which allows us to rewrite the inner product as follows:

\begin{equation}
    \langle \Psi_0^S, \Psi_h^S\rangle=\int_{-\infty}^\infty \frac{d\chi}{\chi^2} \Psi_0^S\left( \chi \right) \Psi_h^S\left( \chi \right)=2\int_{-\infty}^\infty d\chi |\chi|^{2\Delta} \Psi_h^S\left( \chi \right)=2\alpha_0 k^S\left( h \right),
    \label{F0_Psi_A}
\end{equation}
and similarly for the anti-symmetric case.

Bringing everything together and using the formula \ref{F_sum}, we have for the four-point function:

\begin{multline}
    \frac{\mathcal F^{A}\left( \tau_1,\tau_2,\tau_3,\tau_4 \right)}{G\left( \tau_1,\tau_2 \right)G\left( \tau_3,\tau_4 \right)}=
    \alpha_0 \int_0^\infty \frac{ds}{\pi} \frac{s}{\pi \coth \pi s} \frac{k^{A}\left( \frac12+is \right)}{1-k^{A}\left( \frac12+is \right)}\Psi_{\frac12+is}^{A}\left( \chi \right)+\\
    \alpha_0\sum_{n =2j>0}\frac{2j-\frac12}{\pi^2}  \frac{k^{A}\left(2j \right)}{1-k^{A}\left( 2j\right)}\Psi_{2j}^{A}\left( \chi \right),
    \label{F_AA_sum}
\end{multline}
which is of course the same as the sum of all ladders in \cite{Maldacena:2016hyu}, and:
\begin{multline}
   \frac{\mathcal F^{S}\left( \tau_1,\tau_2,\tau_3,\tau_4 \right)}{G\left( \tau_1,\tau_2 \right)G\left( \tau_3,\tau_4 \right)}=
   \alpha_0 \int_0^\infty \frac{ds}{\pi} \frac{s}{\pi \coth \pi s} \frac{k^{S}\left( \frac12+is \right)}{1-k^{S}\left( \frac12+is \right)}\Psi_{\frac12+is}^{S}\left( \chi \right)+\\
   \alpha_0\sum_{n =2j+1>0}\frac{2j+\frac12}{\pi^2}  \frac{k^{S}\left(2j+1 \right)}{1-k^{S}\left( 2j+1\right)}\Psi_{2j+1}^{S}\left( \chi \right).
    \label{F_BB_sum}
\end{multline}

The full four-point function is a combination of the symmetric and antisymmetric parts, and from \ref{odd_even} we find:

\begin{equation}
    \mathcal F = \mathcal F^A - \mathcal F^S.
    \label{F_FA_FS}
\end{equation}

The antisymmetric four-point function (\ref{F_AA_sum}) has a formally divergent contribution from the $h=2$ mode. 
Since we are studying the conformal, or large coupling, limit $\beta J \gg 1$, we can expect that this divergence is present only in this limit and not in the exact solution.
In \cite{Maldacena:2016hyu} it was argued that the kernel receives corrections in $\left( \beta J \right)^{-1}$, which regularize this divergence.

Likewise, the symmetric four-point function (\ref{F_BB_sum}) has a pole in $h$ coming from the $h=1$ mode.
We can hope to regularize it the same way.
In the next section, we consider the $h=1$ contribution in the exactly solvable limit of $q \to \infty$.
We see that at least in this case, the kernel shifted away from one, and the divergence in the four-point function is removed.
Although we do not address the case of general $q$, we expect this regularization to be generic.

\section{\(h=1\) mode}
\label{sec:h=1}
The $U(1)$ charge operator has dimension $h=0$ in the UV.
Since our model respects the $U(1)$ symmetry, the dimension stays the same in the conformal limit as well.
The symmetry of the spectrum under $h \leftrightarrow 1-h$ allows us to identify this $U(1)$ charge with the $h=1$ mode in the symmetric four-point function (\ref{F_BB_sum}).

The charge operator is present in the spectrum of the model in the conformal limit, therefore its dimension solves the equation $k^S\left( \Delta,h \right)=1$.
This means that it produces a pole in the four-point function. 
We have already seen this happen for the $h=2$ mode in the antisymmetric sector, although there is an important difference.
The pole from the $h=2$ mode is present only in the conformal limit, when the coupling is large $\beta J \gg 1$.
This limit possesses full reparameterization symmetry, which gets broken down to $SL(2,\mathbb R)$ symmetry at finite coupling.

The $h=2$ mode corresponds to the generator of this reparameterization symmetry. 
Since the symmetry gets broken at finite coupling, the dimension of the corresponding operator receives corrections of the order of inverse coupling:
\begin{equation}
  h-2\sim \frac{1}{\beta J}.
  \label{h=2+}
\end{equation}
Therefore the eigenvalue of the kernel also gets corrected:
\begin{equation}
  k^A\left( \Delta,h \right) \sim \frac{1}{\beta J},
  \label{k=1+}
\end{equation}
and the pole in the four-point function gets resolved.

At first glance, this resolution cannot happen for the $h=1$ mode. 
Indeed, since $U(1)$ is an exact symmetry of the theory, the dimension of the charge should stay the same regardless of coupling.
Therefore the eigenvalue of the kernel, which is a function of dimension, should not change either. 
However, this is not the case.
The symmetry of the spectrum $h \leftrightarrow 1-h$ is a feature of the conformal limit, and it can be absent in the exact solution of the model.
We consider an exactly solvable limit of $q \to \infty$ to see that the $h=1$ mode indeed receives corrections in the inverse coupling.

\subsection{Correction to \(h=1\) at large \(q\)}
At large $q$, the model simplifies and turns out to be solvable at any value of the coupling \cite{Maldacena:2016hyu}.
We consider the theory at finite temperature $\beta$, with the $\mathcal J$ coupling finite,
\begin{equation}
  \mathcal J\equiv {\sqrt{q} J}2^{1-q/2}.
  \label{J_cal_def}
\end{equation}
and the dimensionless coupling kept large, 
\begin{equation}
  \beta \mathcal J \gg 1.
  \label{bJ>>1}
\end{equation}
The propagator gets a $1/q$ correction which we denote $g(\tau)$:
\begin{equation}
  G\left( \tau \right)=\frac12 \sgn \tau \left( 1+\frac{1}{q}g\left( \tau \right)+O\left( q^{-2} \right) \right), \qquad \Sigma\left( \tau \right)=\frac12 \sgn \tau e^{g\left( \tau \right)}\left( \mathcal J^2 \cdot \frac{1}{q}+O\left( q^{-2} \right) \right),
  \label{G_Sigma_q_inf}
\end{equation}
where $g\left( \tau \right)$ was found in \cite{Maldacena:2016hyu} to be:
\begin{equation}
  e^{\frac{g\left( \tau \right)}2}=\frac{\cos \frac{\pi v}{2}}{\cos \left( \frac{\pi v}{2}-\frac{\pi v}{\beta}|\tau| \right)}, \qquad \beta \mathcal J=\frac{\pi v}{\cos \frac{\pi v}{2}}.
  \label{e^g_sol}
\end{equation}
At large $\beta \mathcal J$, $v \sim 1$. 
So to get a correction in inverse coupling, we consider $v$ near 1 and find the answer as a series in $\left( 1-v \right)$.

The four-point function solves the symmetrized kernel equation (for simplicity, we omit half of the coordinates $\Psi^S$ depends upon):
\begin{equation}
  \int\tilde{K}^S\left( \theta_1,\theta_2|\theta_3,\theta_4 \right)\Psi^S\left( \theta_3, \theta_4\right)=k^S \Psi^S\left( \theta_1,\theta_2 \right). 
  \label{kernel_eq_q}
\end{equation}
The symmetrized kernel is defined as follows:
\begin{equation}
  \tilde{K}^S=|G\left( \theta_{12} \right)|^{\frac{q-2}{2}}K^S\left(\theta_1,\theta_2|\theta_3,\theta_4  \right)|G\left( \theta_{34} \right)|^{\frac{2-q}{2}},
  \label{K_tilde_def}
\end{equation}
and $K^S$ is the finite-temperature version of the kernel (\ref{KA_def}).
Using the propagators (\ref{G_Sigma_q_inf}), we write the eigenvalue equation as:
\begin{equation}
  -\frac{1}{q}\frac{\mathcal J^2}{4}\int d\theta_3 d\theta_4 \sgn\left( \theta_{13} \right) \sgn \left( \theta_{24} \right) e^{\frac12 \left( g\left( \theta_{12} \right)+g\left( \theta_{34} \right) \right)}\Psi^S\left( \theta_3,\theta_4 \right)=k^S \Psi^S\left( \theta_1,\theta_2 \right).
  \label{K_eqn_q_inf}
\end{equation}
As in \cite{Maldacena:2016hyu}, we can apply an operator $\partial_{\theta_1} \partial_{\theta_2} e^{-\frac12 g\left( \theta_{12} \right)}$ to both sides of this integral equation and get a differential equation instead.
Using the ansatz:
\begin{equation}
  \Psi_n^S \left( \theta_1,\theta_2 \right)=\frac{e^{-in \frac{\theta_1+\theta_2}{2}}}{\sin \frac{\tilde{x}}{2}}\psi^S_n\left( \tilde{x} \right), \qquad \tilde{x}\equiv vx + \left( 1-v \right)\pi 
  \label{Psi_ansatz}
\end{equation}
and the eigenvalue of the kernel at large $q$,
\begin{equation}
  k^S \sim \frac{2}{h\left( h-1 \right)}\frac{1}{q}+O\left( q^{-1} \right),
  \label{kS_q_inf}
\end{equation}
we get that the $\psi^S_n$ function satisfies a hypergeometric equation:
\begin{equation}
  \left( \tilde{n}^2+4\partial_{\tilde{x}}^2-\frac{h\left( h-1 \right)}{\sin^2\frac{\tilde{x}}{2}} \right)\psi^S_n\left( \tilde{x} \right)=0, \qquad \tilde{n}\equiv\frac{n}{v}.
  \label{psi_S_eq}
\end{equation}
The solution has to satisfy the correct discrete symmetries.
It has to be anti-periodic in $\theta$,
\begin{equation}
  \Psi^S\left( \theta_1+2\pi, \theta_2 \right)=\Psi^S\left( \theta_1,\theta_2+2\pi \right)=-\Psi^S\left( \theta_1,\theta_2 \right),
  \label{Psi_2pi}
\end{equation}
and symmetric in coordinates:
\begin{equation}
  \Psi^S\left( \theta_1,\theta_2 \right)=\Psi^S\left( \theta_2,\theta_1 \right).
  \label{Psi_S_sym}
\end{equation}
In the first order in $(1-v)$, this translates into discrete symmetries of $\psi^S_n\left( \tilde{x} \right)$ as:
\begin{equation}
  \psi^S_n\left( x+2\pi \right)=\left( -1 \right)^n \psi^S_n\left( x \right), \qquad \psi^S_n\left( -x \right)=-\psi^S_n\left( x \right),
  \label{psi^S_discr}
\end{equation}
or in particular:
\begin{equation}
  \psi_n^S\left( 2\pi -x \right)=\left( -1 \right)^{n+1}\psi^S_n\left( x \right).
  \label{psi^S_2pi_discr}
\end{equation}
The solution to (\ref{psi_S_eq}), satisfying this condition, in the vicinity of zero reads as:
\begin{equation}
  \psi^S_n\left( \tilde{x} \right)\sim \left\{
    \begin{aligned}[r]
      \sin^h \frac{\tilde{x}}{2} \,_2F_1\left( \frac{h-\tilde{n}}{2}, \frac{h+\tilde{n}}{2}; \frac12; \cos^2 \frac{\tilde{x}}{2} \right), \qquad n\text{ odd},\\
    \cos \frac{\tilde{x}}{2}\sin^h \frac{\tilde{x}}{2} \,_2F_1\left( \frac{h-\tilde{n}+1}{2}, \frac{h+\tilde{n}+1}{2}; \frac32; \cos^2 \frac{\tilde{x}}{2} \right),  \qquad n\text{ even}.
  \end{aligned}
\right.
  \label{psi^S_soln}
\end{equation}
To make these functions convergent near zero, we have to make sure that the hypergeometric function truncates into polynomial, that is one of the first two arguments is a non-positive integer.
This can be satisfied if we correct the value of $h$.
In the vicinity of $h=1$, we take:
\begin{equation}
  h_n=1+\left|\tilde{n}\right|-|n|=1+|n|\frac{1-v}{v}+O\left( 1-v \right)=1+\frac{2|n|}{\beta \mathcal J}+O\left( \left( \beta \mathcal J \right)^{-1} \right).
  \label{h_corr}
\end{equation}
Notice that the dimension $h=0$, even with a small correction, does not reduce the eigenfunction to a polynomial. 
This allows us to conclude that the dimension of the charge in the conformal limit is $h=1$ and not $h=0$ as might be expected from the conservation law.

To make the kernel (\ref{kS_q_inf}) meaningful, we also include a $1/q$ correction to $h$:
\begin{equation}
  h_n=  1+\frac{2|n|}{\beta \mathcal J}+\frac{2}{q}+O\left(q^{-1}\right)+O\left( \left( \beta \mathcal J \right)^{-1} \right).
  \label{h_corr_q}
\end{equation}
This shift can be found by considering the next order in $1/q$ in the eigenvalue equation (\ref{kernel_eq_q}).

Then the kernel also gets corrected:
\begin{equation}
  k^S_n=1-\frac{2|n|}{\beta \mathcal J}+O\left( q^{-1} \right)+O\left( \left( \beta \mathcal J \right)^{-1} \right).
  \label{kS_corr}
\end{equation}
Now we can expand the $h=1$ part of the symmetric four-point function in $\Psi^S_n$ eigenfunctions, with the eigenvalue of the kernel given by (\ref{kS_corr}).
The $h=1$ pole in the four-point function gets cured by $\left( \beta \mathcal J \right)^{-1}$ corrections to the kernel.

The correction away from the $q \to \infty$ limit is harder to compute, but we expect it to be corrected by powers of the inverse coupling as well. 
Also, the question of effective action for $U(1)$ symmetry has been discussed in \cite{Fu:2016vas} in the context of $\mathcal N=2$ supersymmetry, where it was found to be non-singular. 

\section{Chaos region}
\label{sec:chaos}

In this chapter, we proceed to find the chaotic exponent for the $h=0$ mode, associated with the $U(1)$ charge. 
A classic result \cite{larkin1969quasiclassical} connects the Lyapunov exponent with a double commutator:

\begin{equation}
    \left\langle \left[ W(t),V(0) \right]^2\right\rangle \sim e^{2\lambda_L t},
    \label{double_comm}
\end{equation}
the intuition being that for conjugated variables $p,q$ the commutator in the semiclassical limit becomes the Poisson bracket, 
\begin{equation}
    \left[ q(t),p(0) \right]\to i\hbar \left\{ q(t),p(0) \right\}=i\hbar \frac{\partial q(t)}{\partial q(0)},
    \label{Poisson_bracket}
\end{equation}
which describes the dependence of the trajectory on the initial conditions. 
Chaos means that trajectories diverge exponentially with time, hence the expected behavior of the double commutator.
However to make this correlator sensible for local operators in a CFT, we need to regularize the correlator  (\ref{double_comm}); to do that, we move one of the commutators halfway down the thermal circle:
\begin{equation}
    \left\langle \left[ W(t),V(0) \right]\left[ W(t+{i\beta}/{2}),V\left( 0+{i\beta}/{2} \right) \right]\right\rangle.
    \label{double_comm_reg}
\end{equation}

All our previous calculations have been done in Euclidean time at zero temperature.
We can formally pass to real time substituting $\tau \to i t$, however in this way we lose information about the ordering of operators.
It is more convenient to describe correlators by making time complex with a small real part,
\begin{equation}
    \tau=it+\epsilon.
    \label{complex_time}
\end{equation}
Going back to our quantum mechanical intuition, a correlator in complex time is:

\begin{equation}
    \left \langle q(\tau)q(0)\right\rangle=\left\langle q\left(i t+\epsilon \right)q(0)\right\rangle=\left\langle q(0)e^{-H\left( it+\epsilon \right)}q(0) \right\rangle, 
    \label{corr_complex}
\end{equation}
converging only for $\epsilon>0$.
Thus we can write a commutator as:
\begin{equation}
    \left[ W(t),V(0) \right]\to W(t)\left( V\left( i\epsilon \right)-V\left( -i\epsilon \right) \right),  
    \label{comm_epsilon}
\end{equation}
implying that the right-hand side is an analytic continuation from Euclidean time.
Bringing everything together, we see that to find the chaos exponent for the complex SYK model, we need the correlator:

\begin{multline}
    \left\langle \bar{\psi}\left( \tau_1 \right)\psi\left( \tau_2 \right)\bar{\psi}\left( t_3 \right)\psi\left( t_4 \right)  \right\rangle=\\
    \left\langle \left( \bar{\psi}\left( \epsilon \right)-\bar{\psi}\left( -\epsilon \right) \right)\psi\left(i t \right)\left( \bar{\psi}\left(\beta/2+ \epsilon \right)-\bar{\psi}\left(\beta/2 -\epsilon \right) \right)\psi\left( \beta/2 + it \right)\right\rangle. 
    \label{psi4_complex_time}
\end{multline}
To compute it, we can apply a similar procedure as before, diagonalizing the retarded kernel instead of the conformal kernel.

\subsection{Retarded kernel}

\begin{figure}
    \centering
    \psfig{file=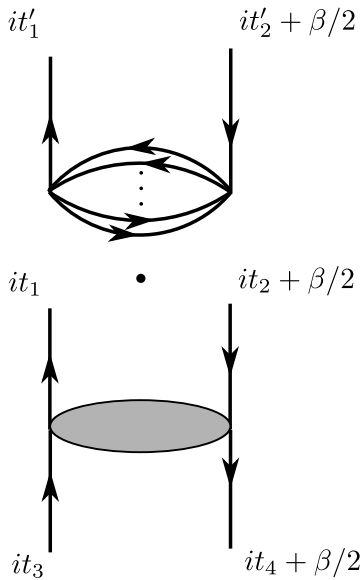, width=.3\textwidth}
    \caption{Retarded kernel acts on a four-point function in complex time.}
    \label{fig:kernel_c_time}
\end{figure}

Retarded kernel is the conformal kernel continued to complex time (see fig.\ref{fig:kernel_c_time}).
Following the prescription of (\ref{psi4_complex_time}), we consider a ladder with one rail at time $it$ and the other at time $\beta/2+it$.
This kernel consist of propagators of two types: ones that go along one rail, and ones which connect one rail to the other.

At finite (unit) temperature, the propagator in complex time is:

\begin{equation}
    G_t\left( \tau_1-\tau_2 \right)=\frac{b \sgn \left( \tau_1-\tau_2 \right)}{|2 \sin \frac12 \left( \tau_1-\tau_2 \right)|^{2\Delta}}\qquad \rightarrow \qquad  G_t\left( \tau_1-\tau_2 \right)=\frac{b \left(\sgn \Re\left( \tau_1-\tau_2 \right)\right)^{2\Delta+1}}{\left(2 \sin \frac12 \left( \tau_1-\tau_2 \right)\right)^{2\Delta}}.
    \label{G_thermal}
\end{equation}

The propagators which go along one rail of the ladder in fig. \ref{fig:kernel_c_time} are conventional retarded propagators:
\begin{multline}
G_R\left( t_1,t_2 \right)=\theta\left( t_1-t_2 \right)\left( G_t \left( it_1+\epsilon,it_2 \right)-G_t\left( it_1-\epsilon ,it_2\right) \right)=\\
    \theta\left( t_1-t_2 \right)\frac{2b \cos \pi \Delta}{\left( 2\sinh \frac12 \left( t_1-t_2 \right) \right)^{2\Delta}}.
    \label{GR_def}
\end{multline}
The propagator connecting left and right rails is as follows:
\begin{equation}
    G_{lr}\left( t_1,t_2 \right)=G\left( it_1+\pi,it_2 \right)=\frac{b}{\left( 2\cosh \frac12 \left( t_1-t_2 \right) \right)^{2\Delta}}.
    \label{Glr_def}
\end{equation}

Just like the conformal kernel, the retarded kernel is diagonalized by even and odd four-point functions, with the eigenvalues differing by a factor of $(q-1)$:
\begin{equation}
    K_R^A=\left( q-1 \right)K_R^S=J^2\left( q-1 \right)G_R\left( t_1,t_2 \right)G_R\left( t_1,t_1' \right)G_R\left( t_2,t_2' \right)G_{lr}\left( t_1',t_2' \right)^{q-2}dt_1'dt_2'.
    \label{KR_def}
\end{equation}
Changing variables to 
\begin{equation}
    z=e^{i\tau},
    \label{z_def}
\end{equation}
we can rewrite the kernel as follows:
\begin{equation}
    K^A_R=\frac{1}{\alpha_0}\theta\left( z_1'-z_1 \right)\theta\left( z_2-z_2' \right)\frac{4 \cos^2 \pi \Delta}{\left| z_1-z_1' \right|^{2\Delta}\left| z_2-z_2' \right|^{2\Delta}\left| z_1'-z_2' \right|^{2-4\Delta}}\left|\frac{z_1z_2}{z_1'z_2'}\right|^{\Delta}dz_1dz_2.
    \label{KR_z}
\end{equation}
We notice that the retarded kernel in this form is very similar to the conformal kernel, so we can easily guess its eigenfunctions:
\begin{equation}
    f_R^S=f_R^A=\frac{\left|z_1'z_2'\right|^\Delta}{\left|z_1'-z_2'\right|^{2\Delta-h}}.
    \label{fR_def}
\end{equation}
The eigenvalue of the antisymmetric retarded kernel is \cite{Maldacena:2016hyu}:
\begin{equation}
    k_R^A\left( h,\Delta \right)=\frac{1}{\pi}\frac{\Gamma\left( -2\Delta \right)}{\Gamma\left( 2\Delta-2 \right)}\Gamma\left( 2\Delta-h \right)\Gamma\left( 2\Delta+h-1 \right)\sin \pi\left( 2\Delta+h \right),
    \label{kRA_ans}
\end{equation}
and the eigenvalue of the symmetric one is the same up to $(q-1)$,

\begin{equation}
    k_R^S\left( h,\Delta \right)=\frac{1}{\pi}\frac{\Gamma\left( 1-2\Delta \right)}{\Gamma\left( 2\Delta-1 \right)}\Gamma\left( 2\Delta-h \right)\Gamma\left( 2\Delta+h-1 \right)\sin \pi\left( 2\Delta+h \right).
    \label{kRB_ans}
\end{equation}
We can rewrite them in terms of conformal kernels:
\begin{equation}
    \frac{k_R^A\left( 1-h \right)}{k^A\left( h \right)}=\frac{\cos \pi \left( \frac{h}{2} -\Delta\right)}{\cos \pi \left(\frac{h}{2}+\Delta \right)}, \qquad \frac{k_R^S\left( 1-h \right)}{k^S\left( h \right)}=\frac{\sin \pi \left( \frac{h}{2} -\Delta\right)}{\sin \pi \left( \frac{h}{2}+\Delta \right)}.
    \label{kR_kC}
\end{equation}
We see that for the series of allowed bound states with integer $h$, the eigenvalues of retarded and conformal kernels are equal:

\begin{equation}
    h \in 2\mathbb Z_+ \qquad \Rightarrow \qquad k_R^A=k^A, \qquad h \in 2\mathbb Z_++1 \qquad \Rightarrow \qquad k_R^S=k^S, 
    \label{k=k_h}
\end{equation}
so both $h=2$ and $h=0$ (or $h=1$) modes develop chaotic behavior.

In general to find the operators contributing to chaos, we solve the equation:
\begin{equation}
    k_R\left( h,\Delta \right)=1.
    \label{kR=1}
\end{equation}
Eigenvalues of the retarded kernel lack the symmetry $h \leftrightarrow 1-h$. 
The solutions of (\ref{kR=1}) for the symmetric kernel consist of the $h=-1$ mode and an infinite series of irrational dimensions:
\begin{equation}
    k_R^S=1\qquad \Rightarrow \qquad h=-1, \qquad {\text{and}} \qquad h=k+2\Delta+O(1/k), \qquad k \in \mathbb Z_+.
    \label{kRA_sol}
\end{equation}
The ``minimal'' case of $q=4$ is an exception: for it, the infinite series is absent and the only mode left is $h=-1$.

Since eigenvalues for odd and even kernels are the same up to $(q-1)$, the solutions to (\ref{kR=1}) for the anti-symmetric kernel consist of a similar infinite series plus the $h=0$ mode:
\begin{equation}
    k_R^A=1\qquad \Rightarrow \qquad h=0, \qquad {\text{and}} \qquad h=k+2\Delta+O(1/k), \qquad k \in \mathbb Z_+,
    \label{kRB_sol}
\end{equation}
again with the exception of $q=4$ case.

To see chaotic behavior, we consider again the three-point function (\ref{fR_def}). 
Substituting back $z_1=e^{-t_1}, z_2=-e^{-t_2}$, we get:

\begin{equation}
    f_R\left( t_1,t_2 \right)=\frac{\exp\left( -\frac{h}{2}\left( t_1+t_2 \right) \right)}{\left( 2\cosh \frac{t_1-t_2}{2} \right)^{2\Delta-h}}.
    \label{fR(t)}
\end{equation}
We see that the three-point function develops exponential growth at $h=-1$; this growth saturates the bound of chaos of \cite{Maldacena:2015waa}.
However, for $h=0$ the Lyapunov exponent in \ref{fR(t)} is zero, hence the $U(1)$ charge does not contribute to chaotic behavior.

\section{Discussion}

In this note we have found the four-point functions for the SYK-like model with complex fermions.
These four-point functions are eigenfunctions of the Casimir of the conformal group which are respectively even and odd under $\chi \to \frac{\chi}{\chi-1}$ symmetry.
To find these eigenfunctions, we have used the shadow formalism.

The shadow formalism naturally allows us to construct eigenfunctions of the Casimir which are odd under time-reversal $\mathcal T$.
Although the usual four-point functions of the SYK-like models are $\mathcal T$-even, at least in the large $N$ limit, it is still an interesting possibility which may be realized for an SYK-like model at a conformal point in the next orders in $1/N$ expansion.

We have also found the eigenvalues of the SYK conformal and retarded kernels for the eigenfunctions of the Casimir.
The eigenvalue equation $k(h)=1$ has a solution $h=1$ corresponding to the $U(1)$ charge operator.
The retarded kernel for this operator is equal to one as well, therefore this operator contributes to chaotic behavior of the model.
However, we find that the corresponding Lyapunov exponent is zero.

The $h=1$ mode creates a divergence in the symmetric four-point function.
A similar divergence is caused by an $h=2$ mode in the usual SYK model with real Majorana fermions.
This divergence is cured by corrections in inverse coupling, when the model is moved away from the conformal limit $\beta J \gg 1$.
We consider the model with $U(1)$ symmetry at large $q$ and arbitrary coupling, and find the corrections to the dimension of the charge operator and the eigenvalue of the kernel in this case.
These corrections regularize the four-point function and make it non-singular. 
We expect the same happen at generic value of $q$.

The $U(1)$ mode is present in the $\mathcal N=2$ SYK model as well, considered in \cite{Fu:2016vas}.
In this case the $U(1)_R$ symmetry is a part of the $\mathcal N=2$ superconformal symmetry, and the charge contributes to the effective action given by an $\mathcal N=2$ super-Schwarzian derivative. 
Therefore one expects the divergence in this mode to be cured by moving away from the (super-)conformal limit, just the way the $h=2$ divergence is removed in the non-supersymmetric real SYK. We return to this question soon in \cite{KB_N2}.

Another interesting question concerns SYK model with gauge symmetries. 
A local symmetry can be sensitive to the reparameterization invariance and its breaking when the theory is moved away from the conformal limit.
We hope to discuss this and related questions elsewhere.

\appendix

\section{Eigenvalues of the kernel}
\label{sec:app_kernel}

We find eigenvalue of the symmetric conformal kernel as an integral (\ref{k_c_comp}):

\begin{multline}
    k^A\left( h \right)=\left.\tau_0^{2h}\int d\tau_1'd\tau_2' K\left( 1,0;\tau_1',\tau_2' \right)f^A_h\left( \tau_1',\tau_2',\tau_0 \right)\right|_{\tau_0 \to \infty}=\\
    \frac{1}{\alpha_0} \int d\tau_1'd\tau_2' \frac{\sgn\left( \tau_1'-1 \right)\sgn \left( \tau_2'\right) }{|\tau_1'-1|^{2\Delta}|\tau_2'|^{2\Delta}|\tau_1'-\tau_2'|^{2-4\Delta}}\frac{\sgn \left( \tau_2'-\tau_1' \right)}{|\tau_1'-\tau_2'|^{2\Delta-h}}.
    \label{kA_int_app}
\end{multline}
Changing variables 
\begin{equation}
    \tau\equiv \tau_1'-1,\qquad \tau'\equiv\frac{\tau_1'\tau_2'}{\tau_2'-1}
    \label{var_change_kernel}
\end{equation}
we find:
\begin{equation}
    k^A\left( h \right)= \frac{1}{\alpha_0} \int d\tau \frac{\sgn \tau}{|\tau|^{2\Delta} |1-\tau|^{1-h}}\int d\tau' \frac{\sgn \tau'}{|\tau'|^{2\Delta}|1-\tau'|^{h}},
    \label{kA_ans_int}
\end{equation}
Note that this expression is symmetric under $h \leftrightarrow 1-h$. 
Using the integral definition of the beta function,
\begin{equation}
    B\left( x,y \right)\equiv \int_0^1 t^{x-1} \left( 1-t \right)^{y-1}dt,
    \label{beta_def}
\end{equation}
we can write it as:
\begin{equation}
    k^A = \frac{1}{\alpha_0}\left( -B\left( 1-2\Delta,2\Delta-h \right)+B\left( h,1-2\Delta \right)+B\left( h,2\Delta-h \right) \right)\cdot \left( h \leftrightarrow 1-h \right),
    \label{kA_c_beta}
\end{equation}
which simplifies to (\ref{kA_ans}).

For the eigenvalue of the anti-symmetric kernel we find:

\begin{equation}
    k^S\left( h \right)=\frac{1}{q-1}\frac{1}{\alpha_0} \int d\tau_1'd\tau_2' \frac{\sgn\left( \tau_1'-1 \right)\sgn \left( \tau_2'\right) }{|\tau_1'-1|^{2\Delta}|\tau_2'|^{2\Delta}|\tau_1'-\tau_2'|^{2-4\Delta}}\frac{1}{|\tau_1'-\tau_2'|^{2\Delta-h}}.
    \label{kB_int_app}
\end{equation}
Making the same change of variables (\ref{var_change_kernel}), we find: 
\begin{equation}
    k^S\left( h \right)=\frac{1}{\alpha_0}\frac{1}{q-1} \int d\tau \frac{\sgn \tau \sgn \left( 1-\tau \right)}{|t|^{2\Delta} |1-\tau|^{1-h}}\int d\tau' \frac{\sgn \tau' \sgn\left( 1-\tau' \right)}{|\tau'|^{2\Delta}|1-\tau'|^{h}},
    \label{kB_int_ans_app}
\end{equation}
which is again written in terms of beta functions:
\begin{equation}
    k^S=\frac{1}{q-1} \frac{1}{\alpha_0}\left( -B\left( 1-2\Delta,2\Delta-h \right)+B\left( h,1-2\Delta \right)-B\left( h,2\Delta-h \right) \right)\cdot \left( h \leftrightarrow 1-h \right),
    \label{kB_beta}
\end{equation}
which simplifies to (\ref{kB_ans}).

The same integrals appear in the inner products with zero-rung four-point functions. 
Writing the inner product with a symmetric function (\ref{F0_Psi_A}),
\begin{equation}
    \langle \Psi_0^S, \Psi^{S}_h \rangle=2\int dy \frac{d\chi}{\chi^2} \frac{|\chi^h|\sgn \chi}{|\chi-y|^h|1-y|^{1-h}|y|^h}{|\chi|^{2\Delta}},
    \label{F0_dy_dchi}
\end{equation}
and changing variables,
\begin{equation}
    y=\frac{1}{\tau}, \qquad \chi=\frac{1}{\tau\tau'},
    \label{var_change_product}
\end{equation}
we arrive at the same integral (\ref{kA_ans_int}), giving:
\begin{equation}
    \langle \Psi_0^S, \Psi^{S}_h \rangle=2\alpha_0 k^S(h).
    \label{ans_A_prod}
\end{equation}
The same procedure gives the inner product with an anti-symmetric function:
\begin{equation}
    \langle \Psi_0^A, \Psi^{A}_h \rangle=2\alpha_0 k^A(h).
    \label{ans_B_prod}
\end{equation}

Similar integrals appear in the computations of retarded kernels. 
The eigenvalue of the antisymmetric kernel is:
\begin{equation}
    k_R^A\left( h \right)=\frac{4\cos^2 \pi \Delta}{\alpha_0}\int\theta\left( -z_1' \right)\theta\left( z_2'-1 \right) dz_1'dz_2'\frac{1}{\left|z_1'\right|^{2\Delta}\left|1-z_2'\right|^{2\Delta}\left|z_1'-z_2'\right|^{2-2\Delta-h}}.
    \label{KRA_int}
\end{equation}
Making a change of variables similar to (\ref{var_change_kernel}):

\begin{equation}
     z_1'=\frac{\left( 1-\tau \right)\tau'}{\tau'-1},\qquad z_2'=1-\tau,
    \label{var_change_kernel_retarded}
\end{equation}
we find that:
\begin{equation}
    k_R^A\left( h \right)=\frac{\sin 2 \pi \Delta}{\pi}\frac{\left( 1-2\Delta \right)\left( 1-\Delta \right)}{\Delta}\int_{-\infty}^0dt \frac{1}{\left|\tau\right|^{2\Delta}\left|1-\tau\right|^{1-h}} \int_0^1 d\tau' \frac{1}{\left|\tau'\right|^{2\Delta}\left|1-\tau'\right|^{h}}.
    \label{KRA_int_ans}
\end{equation}
Employing again the integral form of the beta function, we derive (\ref{kRA_ans}).
The symmetric kernel is also proportional to this expression.

\section{Four-point function}
\label{sec:app_4pt}

The four-point function is given in (\ref{PsiA_ff},\ref{PsiB_ff}) as an integral of a product of three-point functions,

\begin{equation}
    \Psi_h=\int d\tau_0 \frac{f_h\left( \tau_1,\tau_2,\tau_0 \right)f_{1-h}\left( \tau_3,\tau_4,\tau_0 \right)}{G\left( \tau_1,\tau_2 \right)G\left( \tau_3,\tau_4 \right)}.
    \label{Psi_int}
\end{equation}
Rewriting this expression in terms of the cross-ratio $\chi$ and using the ansatz (\ref{3pt_formula}), we have for instance for the symmetric eigenfunction:
\begin{equation}
    \Psi_h^{S}\left( \chi \right)=\int dy \frac{|\chi|^h|1-y|^{h-1}}{|y|^h |\chi-y|^h}.
    \label{Psi_AA}
\end{equation}
To take this integral, we employ the definition of the hypergeometric function:

\begin{equation}
    \,_2F_1\left( a,b;c;\chi \right)=B\left( b,c-b \right)\int_0^1 dxx^{b-1}\left( 1-x \right)^{c-b-1}\left( 1-\chi x \right)^{-a}, \qquad 0<\chi<1,
    \label{2F1_def}
\end{equation}
with $B\left( b,c-b \right)$ being the Euler beta function.
Changing variables and renaming parameters, we can derive a set of analogous identities for $0<\chi<1$ and for $\chi>1$.
The eigenfunctions for $\chi<0$ can be restored from $\chi \to \frac{\chi}{\chi-1}$ symmetry.

Using these identities, it is easy to find the integral (\ref{Psi_AA}) on the four intervals for $0<\chi<1$:

\begin{eqnarray}
    y<0: &\qquad & \frac{1}{2 \cos \pi h}\left( F_h\left( \chi \right)-F_{1-h}\left( \chi \right) \right)=\pi \sin{\pi h} G_h\left( \frac{\chi-1}{\chi} \right),\\
    0<y<\chi: & \qquad & F_{1-h}\left( \chi \right),\\
    \chi<y<1: &\qquad & \frac{1}{2 \cos \pi h}\left( F_h\left( \chi \right)-F_{1-h}\left( \chi \right) \right)=\pi \sin{\pi h} G_h\left( \frac{\chi-1}{\chi} \right),\\
    y>1: & \qquad & F_{h}\left( \chi \right),
    \label{0<chi<1}
\end{eqnarray}
and for $\chi>1$:
\begin{eqnarray}
    y<0:&\qquad&  \frac12 G_h\left(\frac{\chi-1}{\chi} \right),\\
    0<y<1:&\qquad& \frac12 G_h\left(\frac{1}{\chi} \right),\\
    1<y<\chi: &\qquad & \frac12 G_h\left(\frac{\chi-1}{\chi} \right),\\
    y>\chi: &\qquad&  \frac12 G_h\left(\frac{1}{\chi} \right),
    \label{chi>1}
\end{eqnarray}
with $F_h$, $G_h$ defined in (\ref{Fh_def}, \ref{Gh_def}).
Summing these terms up and including the sign functions, we get the even and odd eigenfunctions (\ref{IAA_FG},\ref{IBB_FG}).

\section{\(\mathcal T\)-odd four-point functions}
\label{sec:app_T}

In addition to the symmetric and anti-symmetric eigenfunction, we can also find the functions with mixed symmetries (odd under exchange of one pair of coordinates and even under exchange of the other pair).
From Section \ref{sec:symmetries}, we know that these eigenfunctions break the time-reversal symmetry.
Using the shadow formalism, we can write them as follows:

\begin{eqnarray}
    \Psi_h^{AS}&=& \int d\tau_0 \frac{f_h^A\left( \tau_1,\tau_2,\tau_0 \right)f^S_{1-h}\left( \tau_3,\tau_4,\tau_0 \right)}{G\left( \tau_1,\tau_2 \right)G\left( \tau_3,\tau_4 \right)},\\
    \Psi_h^{SA}&=& \int d\tau_0 \frac{f_h^S\left( \tau_1,\tau_2,\tau_0 \right)f_{1-h}^A\left( \tau_3,\tau_4,\tau_0 \right)}{G\left( \tau_1,\tau_2 \right)G\left( \tau_3,\tau_4 \right)}.
    \label{Psi_mixed_def}
\end{eqnarray}

The first function is even under $\chi \to \frac{\chi}{\chi-1}$ and the second one is odd.
Using the ansatz (\ref{3pt_formula}) for the three-point functions, we find:

\begin{equation}
    \Psi^{AS}_h\left( \chi \right)=\left\{
        \begin{gathered}
            2\sin^2 \frac{\pi h}{2}G_h\left( \frac{1}{\chi} \right), \qquad \chi<0,\\
            2\sin^2 \frac{\pi h}{2}G_h\left( \frac{\chi-1}{\chi} \right), \qquad 0<\chi<1,\\
            0, \qquad \chi>1.
        \end{gathered}
        \right.
    \label{IAB_FG}
\end{equation}

\begin{equation}
    \Psi^{SA}_h\left( \chi \right)=\left\{
        \begin{gathered}
            -2\cos^2 \frac{\pi h}{2}G_h\left( \frac{1}{\chi} \right), \qquad \chi<0,\\
            2\cos^2 \frac{\pi h}{2}G_h\left( \frac{\chi-1}{\chi} \right), \qquad 0<\chi<1,\\
            0, \qquad \chi>1.
        \end{gathered}
        \right.
    \label{IBA_FG}
\end{equation}

As we have seen before from symmetry considerations, the $\mathcal T$-odd four-point functions vanish for $\chi>1$.
Although this function is not continuous at $\chi=1$, it can be checked that the eigenvalue equation for the Casimir does not have a singularity in the right-hand side:

\begin{equation}
    \mathcal C \Psi^{AS}_h - h\left( h-1 \right) \Psi^{AS}_h=0.
    \label{C_mixed}
\end{equation}

One might wonder how it is possible for a Casimir to have four independent eigenfunctions for each eigenvalue.
Indeed, the Casimir is a differential operator of the second order, so for each eigenvalue it should have two distinct eigenfunctions.
The answer is that this statement only holds locally: inside each of the regions $\chi<0$, $0<\chi<1$, $\chi>1$, there are two independent eigenfunctions. 
When $\chi<1$, the $\mathcal T$-breaking eigenfunctions are linear combinations of the $\mathcal T$-preserving ones:

\begin{equation}
    \Psi^{AS}_h\left( \chi \right)=\left\{
        \begin{gathered}
            \sin^2 \frac{\pi h}{2}\left( \Psi^A\left( \chi \right)-\Psi^S\left( \chi \right) \right), \qquad \chi<0,\\
            \sin^2 \frac{\pi h}{2}\left( \Psi^A\left( \chi \right)+\Psi^S\left( \chi \right) \right), \qquad 0<\chi<1,\\
        \end{gathered}
        \right.
    \label{AS_comb}
\end{equation}

\begin{equation}
    \Psi^{SA}_h\left( \chi \right)=\left\{
        \begin{gathered}
            -\cos^2 \frac{\pi h}{2}\left( \Psi^A\left( \chi \right)-\Psi^S\left( \chi \right) \right), \qquad \chi<0,\\
            \cos^2 \frac{\pi h}{2}\left( \Psi^A\left( \chi \right)+\Psi^S\left( \chi \right) \right), \qquad 0<\chi<1.\\
        \end{gathered}
        \right.
    \label{SA_comb}
\end{equation}

When $\chi>1$, the basis of Casimir eigenfunctions consists of two $\mathcal T$-even states. 
On the boundaries of these intervals the eigenfunctions can be smoothly glued together (in the sense that the Casimir equation remains non-singular), forming globally four linearly independent functions.

As we have mentioned before, the melonic limit of the SYK model does not admit $\mathcal T$-odd four-point functions, but they may be relevant for next orders in $1/N$ expansion in the theory with a conformal limit.


\bibliographystyle{JHEP}
\bibliography{complexSYK_writeup1}
\end{document}